\documentclass[11pt]{article}
\usepackage{amsfonts}
\usepackage[dvips]{graphicx}
\begin{document}
\title{\bf{Sharing Graphs}}
\author{K. R. Sahasranand \and Nithin Nagaraj \and \small Department of Electronics and Communication Engineering,\\\small Amrita Vishwa Vidyapeetham, Amritapuri Campus, \\ \small Kollam-690525, Kerala, India.\\{\bf \small Email:~sanandkr@gmail.com, nithin@am.amrita.edu}}
\date{}
\maketitle \abstract \noindent Almost all known secret sharing
schemes work on numbers. Such methods will have difficulty in
sharing graphs since the number of graphs increases exponentially
with the number of nodes. We propose a secret sharing scheme for
graphs where we use graph intersection for reconstructing the secret
which is hidden as a sub graph in the shares. Our method does not
rely on heavy computational operations such as modular
 arithmetic or polynomial interpolation but makes use of very basic operations
 like assignment and checking for equality, and graph intersection can also be
 performed visually. In certain cases, the secret could be reconstructed using
 just pencil and paper by authorised parties but cannot be broken by an adversary
 even with unbounded computational power. The method achieves perfect secrecy for $(2, n)$
 scheme and requires far fewer operations compared to Shamir's algorithm. The
 proposed method could be used to share objects such as matrices, sets, plain text and even a heterogeneous
collection of these. Since we do not require a previously agreed upon
encoding scheme, the method is very suitable for sharing heterogeneous collection of objects in a dynamic fashion.\\


\section{Introduction}

Secrets have to be kept secret. At times, we may choose certain people to be privy to the whole secret. At times, we may also want it in such a way that a secret has to be shared among a number of people and any set with a desired number of people from them can unlock the secret. In secret sharing terminology, this `desired number' is called threshold. If the number of people is $n$ and threshold is $k$, the scheme is called a $(k, n)$ scheme. It also has the property that any number of people less than the threshold $(<k)$ will be unable to reconstruct the secret. The whole problem of secret sharing has drawn considerable attention from researchers; the large variety of secret sharing schemes available today testifies as to this; \cite{shamir}, \cite{mignotte}, \cite{asmuth}, \cite{abhi}, \cite{visual} to name a few.

Shamir's scheme \cite{shamir} and most other secret sharing schemes assume (without stating explicitly) that every secret could be represented as a number and shared according to a threshold scheme. However, we come across such scenarios wherein the secret to be shared is not a number, but an object. Then, encoding the secret to a number and decoding it back upon reconstruction can become a tedious task.

Here, the term object is intended to mean a piece of information manifested in some convenient form. It could be one of (but not limited to) the following: a set, a matrix, a graph or a data structure defined over numbers, characters, bytes or even raw symbols (which may or may not carry any meaning). It can also be a heterogeneous collection of these. For example, a person may keep as secret his username (string of alphabets), account number (an integer), bank balance (a floating point number), password (which in itself may contain numbers, alphabets and special symbols) and a map to some secret destination (which may be a graph) all together. If the person wants to share this heterogeneous secret set, how should he go about doing it?

Consider the case where the secret to be shared is a graph of a fixed number of vertices $n$ ($n$ is publicly known). Then, there are $2^{^nC_2}$ different graphs possible, the secret being one of them. The scenario could be one in which a file is compressed using a Huffman tree. Only with the tree can one decode the compressed file. Then, it is reasonable that the file be publicly available and the Huffman tree, a secret to be shared among a set of people. Large amount of information such as a protein structure or the class diagram of a complex system or the circuit diagram of VLSI may be modelled as graphs (\cite{prote}, \cite{proofbio}, \cite{vlsi}) and there could arise a need for them to be shared according to some threshold secret sharing scheme among a set of people in a convenient fashion. Other examples include cartographic and demographic applications.

In this paper, we propose a new secret sharing algorithm for graphs. The secret graph is treated as a sub graph in each of the shares and reconstruction of the secrets is achieved by intersection of graphs. Graph intersection can also be performed visually and hence our method does not rely on heavy computational operations such as modular arithmetic and polynomial interpolation like conventional secret sharing algorithms. The problem of encoding graphs into numbers is eliminated in our method since we directly deal with graphs for generation of shares and for reconstruction of the secret. We propose a $(2,n)$ threshold scheme and claim that our method is perfectly secure as no information of the secret is revealed by the shares. Computationally, our method outperforms Shamir's scheme for sharing graphs. For sharing passwords of length $8$, one can encode it into a graph with as less as $11$ nodes and use our proposed algorithm. This demonstrates the practical value of the proposed algorithm.

The paper is organized as follows. We discuss the prevalent problems with the existing schemes. In the next section, we introduce the basic idea of our algorithm using a simple example. Then, we go on to explain how the same idea applies to graphs. The following section deals deals the security aspects of the proposed scheme followed by complexity analysis of the algorithm. Thereafter we enumerate the advantages our scheme has over existing schemes. We end by listing the drawbacks of the scheme. Although we propose possible solutions, much of it is left for future research.

\section{Problems with existing schemes}\label{section:problems}

\begin{figure}[ht]
\centering
\includegraphics[scale=0.6]{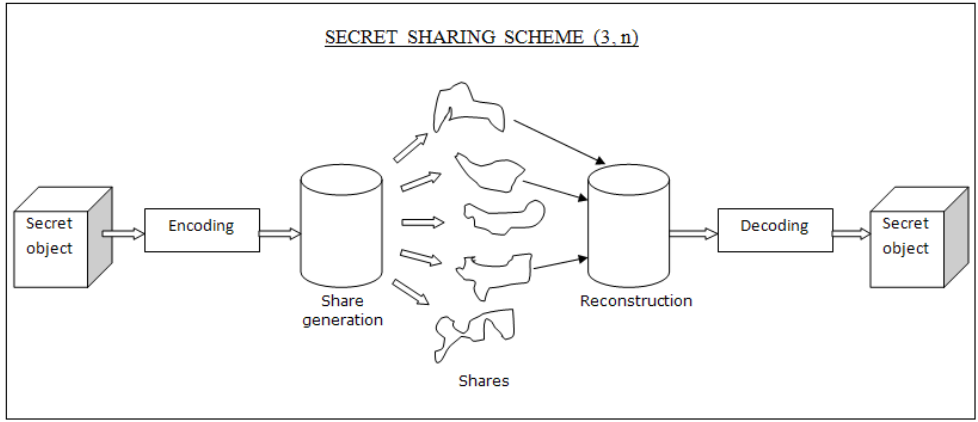}
\caption{Secret Sharing Scheme $(3, n)$}\label{fig:sss}
\end{figure}

First, let us see what it takes to share an object using existing secret sharing schemes. Apart from Visual cryptography \cite{visual}, to the best of our knowledge, all secret sharing schemes share only integers. They work on the assumption that the data to be shared is (or can be made) a number \cite{shamir}. However, they fail to address the task of encoding and decoding (see Figure \ref{fig:sss}), which can be quite tedious in certain scenarios.

The major problems with using existing schemes such as Shamir's \cite{shamir} in scenarios such as those mentioned earlier are enumerated below:

\begin{enumerate}
\item Two extra steps are required, viz., encoding while creating shares and decoding after reconstruction from shares. Let us examine what we mean by encoding, in this context. The secret to be shared could be an object like graph, set etc. During encoding, we are defining a mapping for that object to an integer.

Suppose we have a set of names \{Ram, Jai, Sri, Dev\}. We assign numbers to each of them uniquely, say, Ram = $2$, Jai = $3$, Sri = $6$, and Dev = $8$. Thus, whenever we want to communicate one of these names, we use the corresponding number. The party at the other end of the communication channel looks up the number-name mapping (which we assume is publicly available and not a secret) and retrieves the intended name.

Here, it seems easier with small numbers being used for (comparatively) larger names. However, an encoding of Ram = $1762536123197$, Jai = $1042352443225$, Sri = $3562716934352$ and Dev = $1265460383679$ would be cumbersome. We come across such scenarios where existing methods actually force us to take up such encoding schemes. We shall soon give an example where we have to encode a large number of objects into numbers.
\label{item:prob_enco}

\item Graphs such as those with around $100$ nodes need $4950$ bits for their representation. Let us observe how this is so:

For counting's sake, we consider choosing an edge the same as choosing two nodes, both being equivalent. Thus for an $n$-node graph, there are $^nC_2$ ways by which an edge could be picked. Each of these edges may or may not be part of a graph formed by these $n$ nodes. Thus, there is a total of $2^{^nC_2}$ possible graphs in $n$ nodes. For $n=100$, this number is $2^{4950}$ (which has approximately $1490$ digits). Thus we will be dealing with $4950$ bit integers for representing each of these graphs if at all we are encoding them. Most secret sharing techniques (\cite{shamir}, \cite{abhi}) use prime numbers and have to do complex calculations involving modular arithmetic and polynomial interpolation. Finding primes as large as this is also not a trivial task.

Even for smaller graphs, since the bit requirement grows in a quadratic fashion, $O(n^2)$, it becomes more and more difficult to find large primes as $n$ increases. Also, the increase in the number of bits processed increases the complexity of interpolation and other calculations such as encoding, creating shares, decoding etc.
\label{item:prob_large}

\item In scenarios involving protein structure, such as residue based packing motifs \cite{prote} or protein-protein interaction \cite{ppi} etc., it is reasonable to expect graphs with number of nodes in thousands \cite{proofbio}. Needless to say, dealing with them using the existing methods is quite difficult.
\label{item:prob_prote}

\item Existing schemes are bound to operate within certain constraints. Shamir's scheme requires a finite field $\mathbb{Z}_p$ to work and hence large primes. The scheme in \cite{abhi} is seen to fail in certain cases (illustrated in \cite{notto}) owing to its dependence on polynomial interpolation for implementation. Visual cryptography \cite{visual}, although is devoid of complex calculations and primes, requires a high resolution image of the secret, a device to create shares etc. What is desirable is a scheme which would work with or without the use of computers and does not involve too much of calculations.

\item If we could come up with a scheme to share graphs it would be useful in other ways also. For instance, we all use passwords somewhere or the other. Here we consider $8$-length alphanumeric case sensitive passwords. The total set to choose from is thus composed of $26$ lower case letters, $26$ upper case letters and $10$ digits, summing up to $62$. Therefore, the total number of possible passwords turns out to be $62^8$, which is approximately $2^{47.6}$. Thus, rather than encoding the passwords into integers and working on primes as large as this, we can encode them into graphs with $11$ nodes. Doing the calculation in previous lines shows that this accounts for $2^{55}$ different graphs, hence $55$ bits. Here, although we have to use $55$ bits where there is only a requirement for $48$ bits, we do have a number of advantages, not having to deal with large primes being one of them. We treat the $55$ bit number as a stream of bits (meaningful when interpreted as a graph) rather than a whole number. This helps to avoid the need to deal with large primes. We enumerate the advantages in a later section.\label{item:prob_passwd}

\end{enumerate}

\section{An algorithm using set intersection}\label{section:set}

\noindent We illustrate a simple yet effective secret sharing algorithm using set operation.

Suppose the secret to be shared is a set of integers, $S = \{0, 2, 13\}$. $S$ is to be shared among $5$ people in such a way that if any $3$ of them come together, $S$ could be reconstructed. It is thus a $(3, 5)$ scheme. We intend to achieve this using set intersection. Thus, the shares are created as:\\

\noindent Share $s_1 = \{-48, -25, -18, -5, 0, 1, 2, 9, 10, 13, 19, 24, 40, 52, 88\}$\\
Share $s_2 = \{-92, -48, -18, -3, 0, 2, 3, 4, 10, 11, 12, 13, 37, 61, 90\}$\\
Share $s_3 = \{-75, -53, -44, -25, -10, -3, 0, 1, 2, 11, 13, 40, 46, 58, 61\}$\\
Share $s_4 = \{-81, -75, -44, -10, -5, 0, 2, 3, 12, 13, 23, 24, 50, 52, 90\}$\\
Share $s_5 = \{-92, -81, -53, 0, 2, 4, 9, 13, 19, 23, 37, 46, 50, 58, 88\}$\\

We shall soon explain how these shares are created. The secret set, S could be reconstructed by performing set intersection operation on three or more of the shares. Intersection here means to pick the common elements. To take the intersection of two sets, say, A and B, is to choose only those elements which are in A as well as in B. For intersection of three sets, A, B and C, we may either

\begin{enumerate}
\item Pick all those elements which are in A as well as B as well as C, or,
\item Take intersection of A and B and then intersect this result with C.
\end{enumerate}

The same applies for higher number of set intersections as well. Here, for reconstructing $S$, we need to intersect any three of the shares, since it is a $(3, 5)$ scheme. However, intersection of two shares does not reveal $S$. For example, intersection of shares $s_2$ and $s_3$ results in a set, $\{-3, 0, 2, 11, 13, 61\}$. Only when this set is intersected with any one of the other three shares do we get $S$.

Clearly, it is possible to perform reconstruction of the secret even without the use of any computing device. It could easily be carried out by a layman visually; may be called \emph{pencil and paper} secret sharing. However, it can be done using a computer algorithm as well. Now we formally describe how shares are created for a $(3, n)$ scheme. We assume that the cardinality of the secret set to be shared is publicly known and is taken to be $u$. We also assume a source that gives distinct random values.\\

\noindent \emph{Algorithm for creating shares}
\begin{enumerate}
\item Create $n$ sets (which are going to be the shares), add the elements in the secret to all of these and mark the elements.
\item For each share $s_i$, $1 \le i \le (n-1)$, do the following:
\begin{enumerate}
\item Add $(n-i).u$ elements at random to the share $s_i$.
\item For each share $s_j$, $j > i$, do:
\begin{enumerate}
\item Randomly pick $u$ unmarked elements from share $s_i$ and mark them.
\item Add these $u$ elements to $s_j$ and mark them.
\end{enumerate}
\end{enumerate}
\item STOP.
\end{enumerate}

\noindent \\Let us see how shares were created for the secret set $S = \{0, 2, 13\}$ using the algorithm for a $(3,5)$ scheme.\\

$n = 5, u = 3$.
\begin{enumerate}
\item $5$ shares, $s_1, s_2, s_3, s_4, s_5$ each containing $S$, are created.
\item To the set $s_1$, $(n-i).u = 12$ random values $(-48, -25, -18, -5, 1, 9, 10, 19, 24, 40, 52, 88)$ are added.
\item $u$ distinct elements from this set are added to each of the shares $s_2, s_3, s_4, s_5$. \\
$(-48, -18, 10)$ are added to $s_2$.\\
$(-25, 1, 40)$ are added to $s_3$.\\
$(-5, 24, 52)$ are added to $s_4$.\\
$(9, 19, 88)$ are added to $s_5$.
\item To the set $s_2$, $(n-i).u = 9$ random values $(-92, -3, 3, 4, 11, 12, 37, 61, 90)$ are added.
\item $u$ distinct elements from this set are added to each of the shares $s_3, s_4, s_5$. \\
$(-3, 11, 61)$ are added to $s_3$.\\
$(3, 12, 90)$ are added to $s_4$.\\
$(-92, 4, 37)$ are added to $s_5$.
\item To the set $s_3$, $(n-i).u = 6$ random values $(-75, -53, -44, -10, 46, 58)$ are added.
\item $u$ distinct elements from this set are added to each of the shares $s_4, s_5$. \\
$(-75, -44, -10)$ are added to $s_4$.\\
$(-53, 46, 58)$ are added to $s_5$.
\item To the set $s_4$, $(n-i).u = 3$ random values $(-81, 23, 50)$ are added.
\item $u$ distinct elements from this set are added to share $s_5$. \\
$(-81, 23, 50)$ are added to $s_5$.
\item To the set $s_5$, nothing more is added since $(n-i).u = 0$.
\end{enumerate}

Now, all the shares have $n.u = 15$ elements each.

\subsection*{Drawbacks of set intersection algorithm}
The set intersection is not perfectly secure. In this context,
perfect secrecy means to ensure that all possible secrets could be
constructed out of a given share. The shares created in the set intersection 
algorithm give away information about what all secrets are possible 
and what all are not. For example, by looking at the share $s_1$, 
one can say that the set $\{1,2,3 \}$ is not the secret.

\section{A new algorithm for sharing graphs}

\noindent For sharing graphs, the same algorithm could be modified
to attain a \emph{perfectly secure} scheme. We shall propose a (2,
n) scheme here. The algorithm for creating shares is described as
follows:

First we create a single share from the secret and copy it $n$ times and modify each share a little.\\

{\bf Algorithm:} \emph{create share}()

\begin{enumerate}
\item Read the secret graph to be shared; it has $c$ nodes.
\item Add $b$ extra nodes to the existing secret graph on $c$ nodes.
\item Connect them to the graph as well as among themselves in a random fashion.
\item Create $n$ copies of the resulting graph and relabel the
 nodes of each of the shares in such a way that the nodes which
 are part of the secret get the same label and all others different. It should also be ensured that the nodes which are part of the secret sub graph get labels in the same lexicographic order as in the secret. (Same lexicographic order means, for example, a graph with nodes $a, b, c$ may be labelled $12, 25, 52$ respectively but not $12, 52, 25$ respectively).
\item If the list $L_e$ contains the list of edges in the secret sub graph in each share, for each share do the following:
\begin{enumerate}
\item Randomly decide whether to skip the next (sub)step or not.
\item Add new edges, which are not in $L_e$, to the secret sub graph and append this new list of edges to $L_e$.
\item Randomly pick $c$ nodes. If it contains $a$ nodes which are part of the secret,
\begin{enumerate}
\item If the $a$ nodes are not all mutually connected, pick new $a$ nodes.
\item Else, if at least one of all possible edges between the $a$ nodes are present in $L_e$, pick new $a$ nodes.
\item Else, continue.
\end{enumerate}
\item Create a complete graph on the $c$ nodes thus picked.
\end{enumerate}
\item STOP.
\end{enumerate}

The scheme works basically by mixing additional irrelevant data from
the same domain as the secret, with the secret to be shared. Now,
let us see how reconstruction of the secret graph takes place.\\

{\bf Algorithm:} \emph{reconstruct}()
\begin{enumerate}
\item Input $2$ shares which are graphs on $b+c$ nodes.
\item Perform set intersection of nodes (sub routine for set
intersection follows) to get a graph on $c$ nodes.
\item Check for edges common to each of the shares and construct the $c$ node graph.
\item Relabel the resulting graph so as to invert step $4$ in \emph{create share}().
For example, the labels of resulting graph $12, 25$ and $52$
correspond to $a, b$ and $c$ respectively.\\
\end{enumerate}

\noindent Intersection of nodes can be described by the following algorithm:

Let the node set of the two shares for intersection be represented as arrays, $A$ and $B$ and the result be added to $C$. We assume that the arrays are sorted. This is a reasonable assumption to make since the set containing the nodes could be created out of the adjacency matrix representing the graph.\\

{\bf Algorithm:} \emph{set intersection}()

\begin{enumerate}
\item $i=1, j=1, k=1$.
\item Until $i \le N$ as well as $j \le N$ holds, do the following:\label{iter}
\begin{enumerate}
\item If $A[i] = B[j]$, do:
\begin{enumerate}
\item Add $A[i]$ to $C[k]$.
\item Increment $i, j, k$.
\end{enumerate}
\item Else if $A[i] < B[j]$, increment $i$.
\item Else if $A[i] > B[j]$, increment $j$. 
\end{enumerate}
\item STOP.
\end{enumerate}

\begin{figure}[ht]
\centering
\includegraphics[scale=0.8]{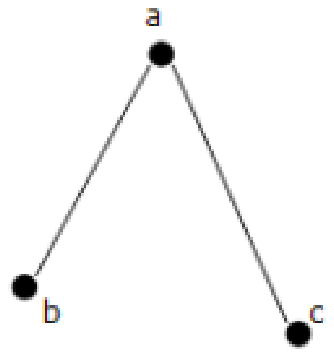}
\includegraphics[scale=0.8]{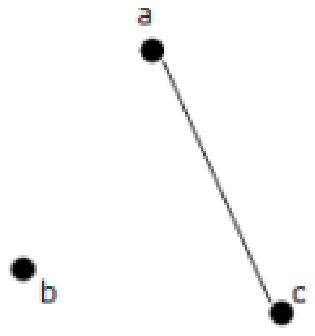}\\
\includegraphics[scale=0.8]{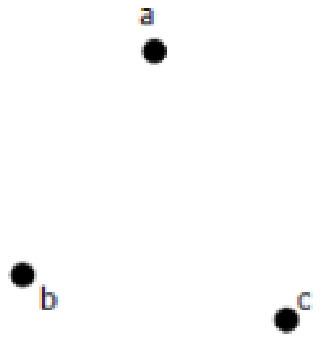}
\includegraphics[scale=0.8]{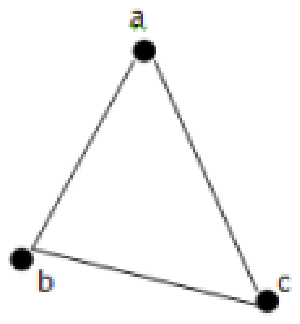}
\hspace{0.1in} \caption{Possible graphs on $3$ nodes without taking
into account, the labels. With labels, the number is
8.}\label{fig:3ng}
\end{figure}

\begin{figure}[tp]
\centering
\includegraphics[scale=0.8]{sec2e.eps}\\
\includegraphics[scale=0.42]{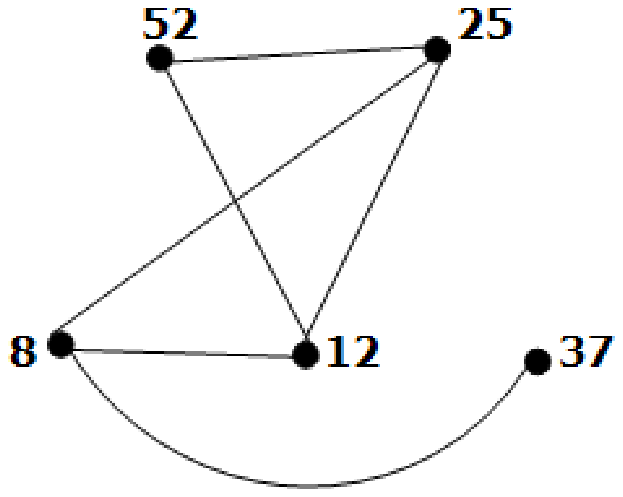}
\includegraphics[scale=0.42]{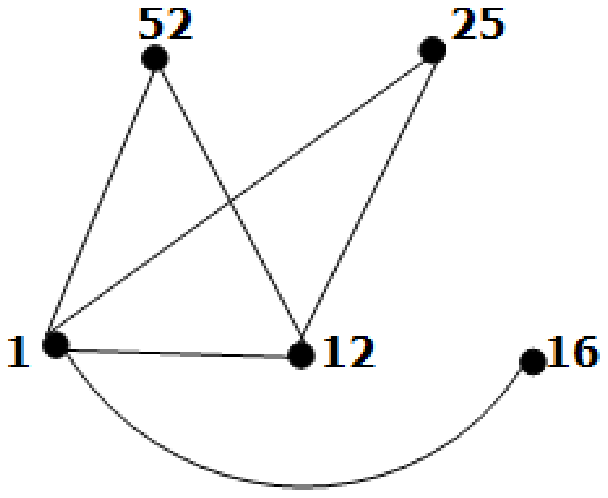}
\includegraphics[scale=0.42]{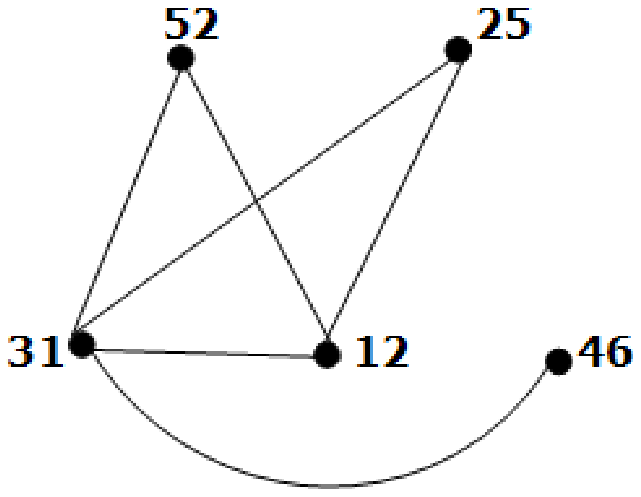}\\
 (a) Above: A 3 node secret. Below: 3 shares for the secret. \\
\includegraphics[scale=0.8]{sec1e.eps}\\
\includegraphics[scale=0.42]{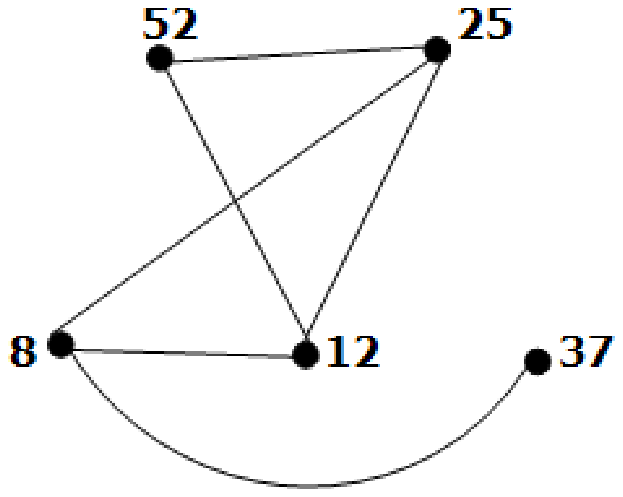}
\includegraphics[scale=0.42]{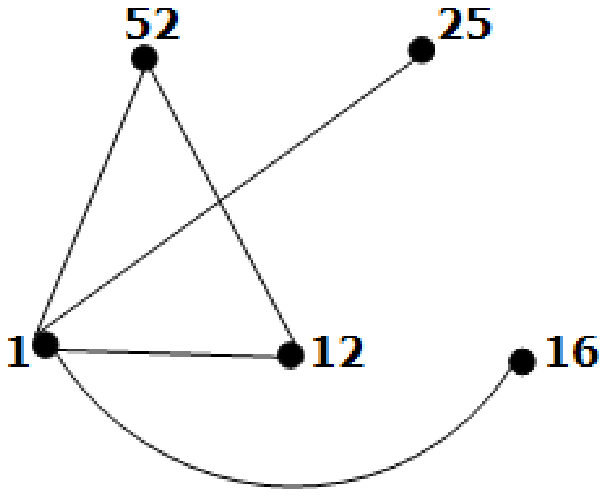}
\includegraphics[scale=0.42]{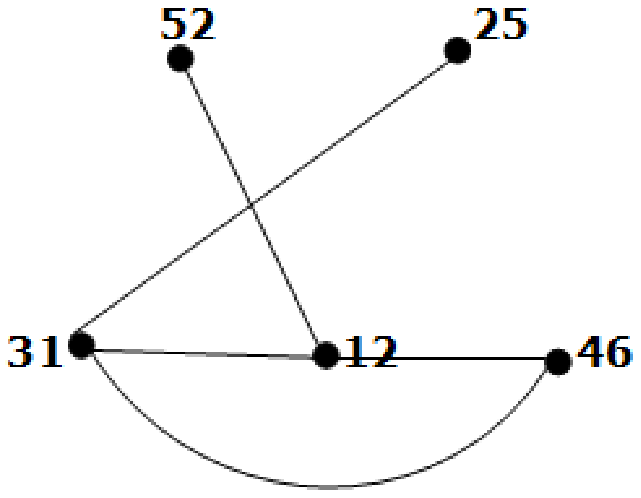}\\
 (b) Above: A 3 node secret. Below: 3 shares for the secret. \\
\caption{$3$ node secrets with $2$ edges and $1$ edge and possible
set of shares for $(2, 3)$ scheme}\label{fig:enum21}
\end{figure}

\begin{figure}[tp]
\centering
\includegraphics[scale=0.8]{sec0e.eps}\\
\includegraphics[scale=0.42]{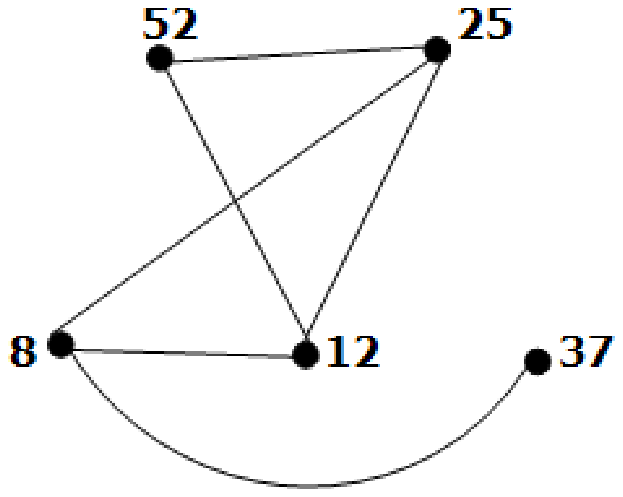}
\includegraphics[scale=0.42]{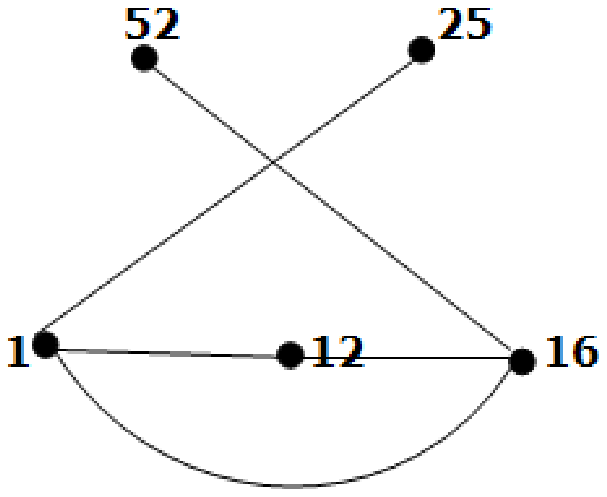}
\includegraphics[scale=0.42]{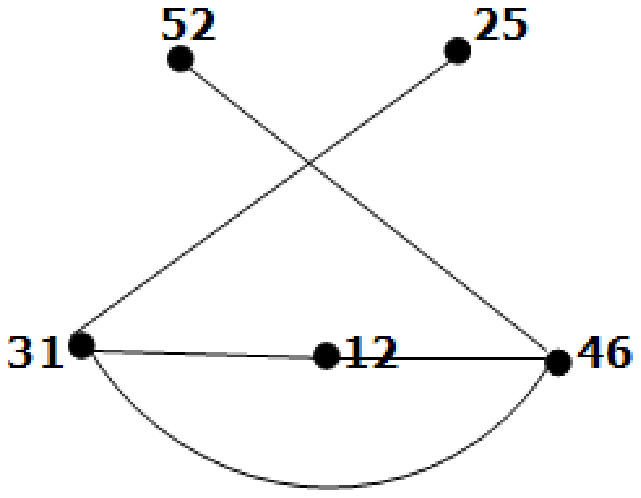}\\
 (a) Above: A 3 node secret. Below: 3 shares for the secret. \\
\includegraphics[scale=0.8]{sec3e.eps}\\
\includegraphics[scale=0.42]{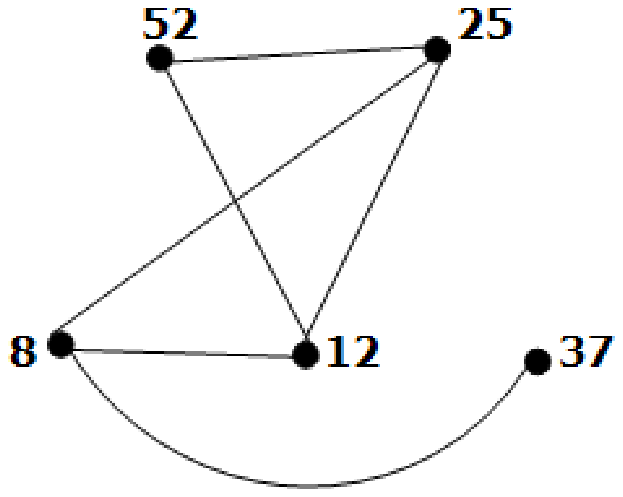}
\includegraphics[scale=0.42]{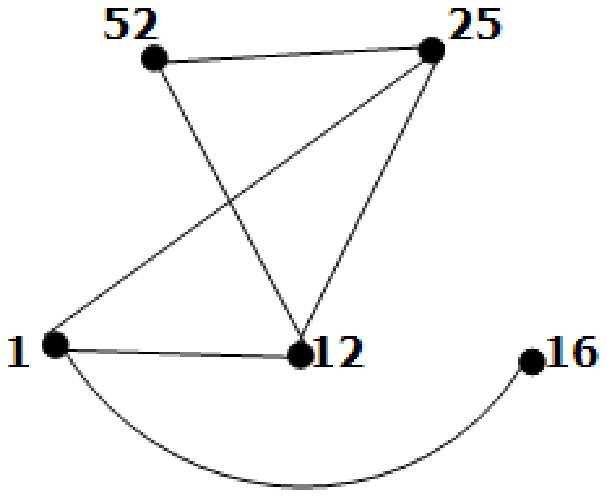}
\includegraphics[scale=0.42]{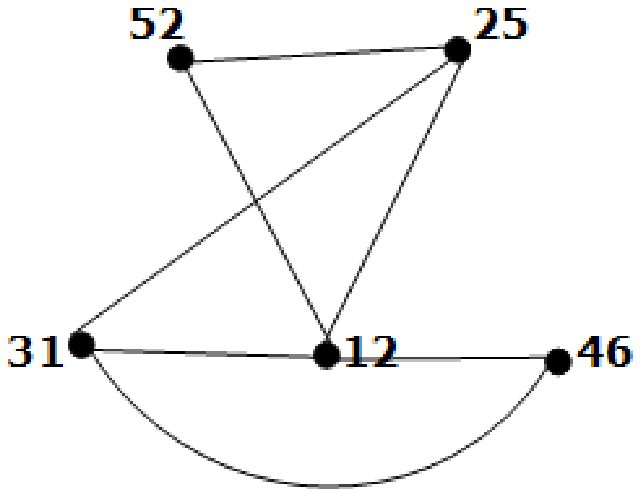}\\
 (b) Above: A 3 node secret. Below: 3 shares for the secret. \\
\caption{$3$ node secrets with no edge and $3$ edges and possible
set of shares for $(2, 3)$ scheme}\label{fig:enum03}
\end{figure}

Figure \ref{fig:3ng} enumerates possible secrets on $3$ nodes, one each with $0, 1, 2$ and $3$ edges. However, in practice, we distinguish between the three possible graphs with $2$ edges as well as the three possible graphs with $1$ edge. The shares generated for these secrets for a $(2, 3)$ scheme are also shown in Figures \ref{fig:enum21} and \ref{fig:enum03}.

\begin{figure}[ht]
\includegraphics[width=0.5\textwidth]{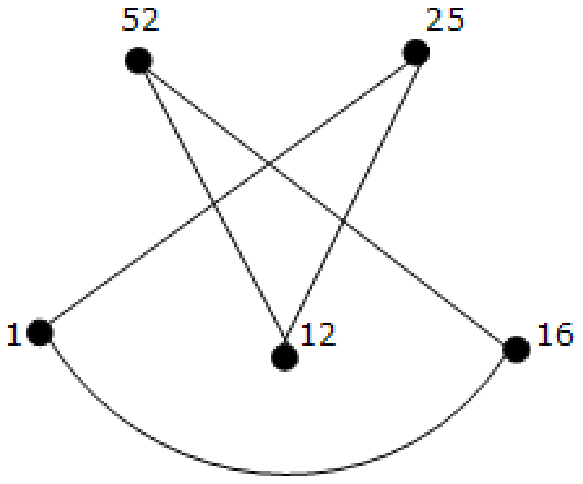}
\includegraphics[width=0.5\textwidth]{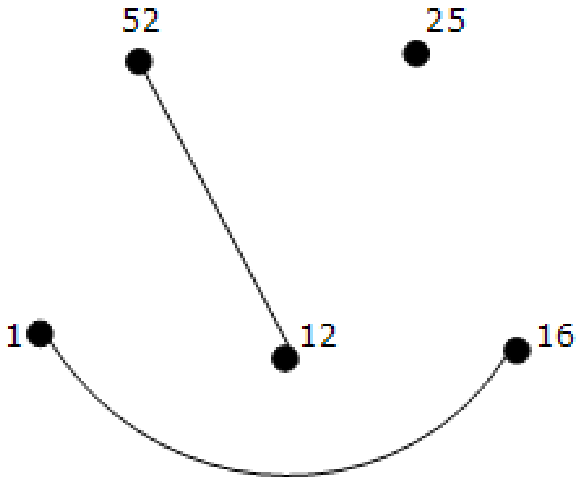}
\caption{Shares which gives away information}\label{fig:counter}
\end{figure}


\subsection*{Proof of Security for a $(2, n)$ scheme}

In Shannon's seminal paper \cite{shannon}, perfect secrecy is defined by a necessary and sufficient condition that, for all cryptograms $E$ the \emph{a posteriori} probabilities are equal to the \emph{a priori} probabilities independently of the values of these.

\noindent To put it mathematically,\\

$P_E(M) = P(M)$ \\

\noindent where, $P_E(M)$ is the \emph{a posteriori} probability of message $M$ if cryptogram $E$ is intercepted and,
$P(M)$ the \emph{a priori} probability of message $M$.

An ideal $(k, n)$ secret sharing scheme is one in which no information about the secret is gained from one share or more than one share upto $k-1$ shares. Such a scheme is said to be perfectly secure. Adopting Shannon's definition to secret sharing, perfect secrecy could be said to have been achieved in a $(2, n)$ scheme if and only if,\\

$P_{S}(D_0) = P(D_0)$ \\

\noindent where, $P_{S}(D_0)$ is the probability of $D_0$ being the secret given a single share $S$ (\emph{a posteriori} probability) and, $P(D_0)$ the \emph{a priori} probability of $D_0$ being the secret.

Now, let us observe how our algorithm for sharing graphs under a $(2, n)$ scheme provides perfect secrecy. Assume that the secret to be shared is an $c$-node graph, $G_c$. Then, the \emph{a priori} probability of $G_c$ is given by:\\

$P(G_c)  = \frac{1}{\emph{Total no. of c-node graphs possible}}$\\

$P(G_c)  = \frac{1}{2^{^cC_2}}$\\

The shares are constructed in such a way that (see Algorithm \emph{create share}) all the shares contain within them a complete graph $C_c$ on $c$ nodes. Let $E_g$ be the edge-set of the secret graph $G_c$, the edge-set of the share be $E_s$ and $E_c$ the edge-set of the complete graph $C_c$. Then,\\

$E_s  = E_c + E_r$ \\

\noindent where $E_r$ is the set of edges in the share excluding the complete graph $C_c$.\\

$E_s  = E_g + (E_c - E_g) + E_r$.\\

$E_c$ could be viewed as $E_g + (E_c - E_g)$ since the edge set of a graph on $c$-nodes is always a subset (may be proper or not) of a complete graph on $c$-nodes.

Thus, a given share could be seen as a decomposition into any secret(on $c$ nodes) and the remaining edges in all possible ways. i.e.,\\

$|E_s|  = |E_g| + (|E_c| - |E_g|) + |E_r|$ \\

\noindent where $|A|$ stands for the cardinality of the set $A$.\\

\noindent $|E_s|  = 0 + (2^{^cC_2} - 0) + |E_r|$. Here the secret is the $c$-node graph with no edge.\\
\noindent $|E_s|  = 1 + (2^{^cC_2} - 1) + |E_r|$. Here the secret is the $c$-node graph with $1$ edge.\\
\vdots
\vdots\\
\noindent $|E_s|  = 2^{^cC_2} + (2^{^cC_2} - 2^{^cC_2}) + |E_r|$. Here the secret is the complete $c$-node graph.\\

Thus, given a single share, it is equally possible that it holds within it any of the possible $2^{^cC_2}$ $c$-node graphs. In other words, keeping one share constant and by varying the second share it is possible to generate all possible secrets on $c$ nodes. This is by virtue of construction of shares. During construction (Algorithm \emph{create share}), edges were added irrespective of the graph we are sharing as a secret. Thus, the number of times each possible sub graph (on $c$ nodes) occurs in the share has nothing to do with the probability of that sub graph being the secret. Therefore, \emph{a posteriori} probability of the secret $G_c$ given a share $S$ is given by:\\ 

$P_S(G_c) = \frac{1}{2^{^cC_2}}$\\

Since, the \emph{a posteriori} probability equals the \emph{a priori} probability, the proposed $(2, n)$ scheme is perfectly secure. Note that the secret need not always be present inside the complete sub graph itself. It may be hidden elsewhere also. This is decided randomly in the algorithm. However, the presence of complete sub graph ensures the possibility of all secrets \emph{at least} once.

\subsection*{Security Analysis}

Here, we analyze the security aspects of the proposed algorithm including an attempt to break it. In this context, breaking means to try to gain some knowledge about the secret given a particular share (since less than $k$ means only one share, for a $(2, n)$ scheme). We will show that the opponent gains absolutely no information about the secret by analyzing a share.

Let us first consider the case of a secret with $3$ nodes. So, there are $2^{^3C_2} = 8$ choices for the secret. Thus, the brute force search involves going through all the $8$ choices.

Now, to share the secret, we apply the algorithm with $b = 2$. i.e., we add $2$ nodes to the secret to create shares. We assume that the number of nodes in the secret is publicly known. Thus, brute force involves choosing $3$ nodes from the $5$ node graph which is a share. This amounts to a total of $^5C_3 = 10$ choices. Clearly, this is greater than the brute force search.

One other possible mode of attack is to analyze the share with respect to edges rather than nodes. Then, it can so result that some graphs are not possible to be the secret, given a share. Figure \ref{fig:counter} shows two such shares (of different secrets). Given that the secret is a $3$-node graph, looking at the share on the left, one can be sure that the complete graph on $3$ nodes is not the secret, because there is no complete sub graph in it and that the share on the right is not a share for either a complete graph or a graph on $2$ nodes, because it has neither as sub graphs. It is to circumvent such a problem that we included the two important steps in the algorithm for creating shares, namely,
\begin{itemize}
\item Adding edges to the secret sub graph
\item Ensuring the presence of a complete graph
\end{itemize}

Thus, over the whole set of choices of $3$ nodes taken in the manner described earlier, the opponent finds that all possible graphs on $3$ nodes including the complete graph could be the secret. i.e., all possible $3$-node graphs are present somewhere in each of the $5$-node share that all of them are equally likely to be the secret. This is because the complete graph can generate all possible graphs on $3$ nodes upon intersection with another share. Thus, the opponent gains absolutely no information about the secret.

\begin{figure}[ht]
\centering \includegraphics[width=0.5\textwidth]{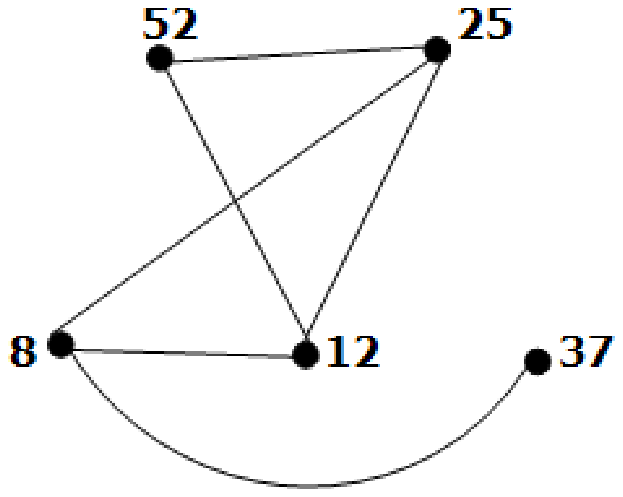}
\caption{A graph which is a share for all secret graphs on $3$ nodes
shown}\label{fig:same}
\end{figure}

There could be a share like the one in Figure \ref{fig:same} in all the cases. i.e., given this share it is possible that it be generated from any of the possible $3$-node graphs. Although this is true for all the shares, we created the same share for all secrets deliberately so that our claim becomes obvious enough.

So, from the example above, it can be seen that perfect security is provided by ensuring that there is a complete graph on $3$ nodes within each secret and choosing the number of nodes to be added in such a way that $^5C_3 \ge 2^{^3C_2}$.
In general, if the secret has c nodes and b nodes are added to it while creating the share, to thwart brute force attack, the necessary condition is that: 
\begin{equation}
^{b+c}C_c \ge 2^{^cC_2}\label{eq:num}
\end{equation}

Thus, the number of nodes to be added (the value $b$) is determined by the above expression. Note that we claim perfect security for $(2,n)$ scheme for graphs because $k-1 = 1$ and information gained from $k-1$ shares is equal to the information gained from one share.

For still larger graphs with $10000$ nodes or so, the search space grows sufficiently large that it provides computational security with a few nodes added although not perfect security. However, note that we cannot do a direct comparison between the search space presented by our scheme and brute force search space to calculate the number of nodes to be added like we did for $3$ node graphs, since the total number of possible graphs is of the order of $2^{49995000}$.

\subsection*{Complexity Analysis}
The scheme we propose makes reconstruction of the secret less complex instead of share generation. Since share generation is a one-time process, it is reasonable that it be complex in lieu of reconstruction which happens more frequently. However, the whole process of creating shares is completed in polynomial time only. Reconstruction is easier still; of the order of $N$($N$, number of nodes in the share).
Let $N = b + c$ and number of shares to be created, $n$.
\begin{enumerate}
\item \emph{Adding nodes}: This step of adding nodes is of the order of number of nodes $N$.
\item \emph{Adding edges}: Adding edges is of the order of $N^2$. i.e., $O(N^2)$.
\item \emph{Relabeling}: Relabeling is of the order of number of nodes $N$, times the number of shares $n$. i.e., $O(N^2)$.
\item \emph{Adding a complete sub graph}: Creating a complete sub graph is $O(N^2)$. This has to be performed for each of the $n$ shares. Thus, the whole step is of the order of $O(N^3)$.
\end{enumerate}

Thus, it is clear that share generation can be carried out in polynomial time. Reconstruction of secret involves set intersection which can be done in $O(N)$ and edge intersection of the resulting set of secret nodes, $O(c^2)$. Since the number of nodes in the secret is much smaller than $N$ always (by Equation \ref{eq:num}), the algorithm for reconstruction could be considered to run in $O(N)$.

However, complexity of attack turns out to be of factorial order. If a share has $N$ nodes and secret is known to have $c$ nodes, picking the possible secret nodes itself is $O(N!)$. This is because such a picking could be done in ${^NC_c}$ ways. Thus, clearly an attack is far difficult compared to share generation and reconstruction, in terms of complexity as well.

Initially, we have just the number of nodes in the secret $c$ to begin with. But, we have defined complexity in terms of $N$, to which $c$ is related in an exponential manner. However, the above analysis serves to compare the complexities of share generation, reconstruction and attack. In the next section, we observe how our scheme is better than one of the conventional schemes in terms of computational complexity.

\subsection*{A comparison with Shamir's scheme}

Let us take a closer look at the complexity using the example of
$8$-length alphanumeric passwords discussed in an earlier section
(Section \ref{section:problems}). Total number of possible passwords
turns out to be $62^8$, which is approximately $2^{47.6}$.\\

{\bf Using graph algorithm:} We can encode the $2^{47.6}$ passwords
into graphs with $11$ nodes since the possible no. of graphs on $11$
nodes is $2^{55}$. We have to use Equation \ref{eq:num} for
calculating the number of nodes to be added. However, since the
brute force search space for passwords is $2^{47.6}$, we substitute
the same in the $RHS$ of the equation and get that we have to add
$93$ nodes so that $^{104}C_{11} \ge 2^{47.6}$. For a $(2,n)$
scheme, assigning new labels is done a total of $93.n + 11$ times
(different labels for the newly added nodes and same label for nodes
in the secret). Picking $11$ nodes and completing the subgraph
requires $55$ operations for one share and $55.n$ for $n$ shares.
Reconstruction can happen in just $104 + 55=159$ steps ($104$ for
intersection of nodes and $55$ for checking intersection of all
possible edges on $11$
node graph just resulted).\\

{\bf Using Shamir's scheme:} For a $(2,n)$ scheme, Shamir's
scheme~\cite{shamir} involves solving a system with $2$ linear
equations and $2$ variables. For example, say, the $2$ shares are
$(x_i, q(x_i))$ and $(x_j, q(x_j))$, then:

\begin{eqnarray*}
& q(x_i) = a_0 + a_1x_i ~~~(mod p),\\ & q(x_j) = a_0 + a_1x_j ~~~(mod p).\\
\end{eqnarray*}

In matrix form, the solution is given as:

\begin{eqnarray*}
\left[
  \begin{array}{c}
    a_0 \\
    a_1 \\
  \end{array}
\right] = \left[
  \begin{array}{cc}
    1 &  x_i\\
    1 & x_j \\
  \end{array}
\right]^{-1} \cdot \left[
  \begin{array}{c}
    q(x_i) \\
    q(x_j) \\
  \end{array}
\right]~~~(mod p),
\end{eqnarray*}

where $a_0$ is the secret. Thus, reconstruction of the secret
involves taking multiplicative inverse over a finite field which is
$O(n^3)$ where $n$ is the no. of bits which in this case is
$48$. 

Thus it can be seen that our algorithm requires 159 basic operations
as opposed to Shamir's which approximately requires $48^3=110592$
basic operations on an average. Our method relies on two basic
operations namely assignment and checking for equality. But,
Shamir's scheme requires addition and multiplication over a prime
modulus for creation of shares and addition, multiplication and
taking inverse over the prime modulus for reconstruction. This
explains why our algorithm is much faster than Shamir's.

\section{Advantages over conventional schemes}
\begin{enumerate}
\item Since our method only relies on intersection of sets, there is no requirement for a previously agreed upon encoding scheme among the parties.  Existing secret sharing schemes in literature invariably assume that the parties who are sharing the secret object have agreed upon a common encoding scheme. Our method circumvents this problem. This enables enormous flexibility for sharing heterogeneous collection of objects in a dynamic fashion.
\item Although encoding graphs into numbers may be considered a triviality, every digital information being already encoded, one advantage our method offers is that we do not have to deal with the resulting bit stream as whole number(s). In our method, the stream of bits needs to be interpreted only as a graph in order for it to be shared as a secret unlike in conventional schemes wherein the stream of bits resulting out of encoding has to be treated as a whole number or a set of whole numbers (in case the bit stream is chopped down).
\item Our method does not require prime numbers for implementation.
\item Unlike in \cite{shamir}, \cite{mignotte}, \cite{asmuth}, \cite{abhi}, no complex operations such as polynomial interpolation, matrix multiplication, solving a system of linear equations or those involving cumbersome modular arithmetic such as finding multiplicative inverses are required.
\item Our reconstruction method does not require computational resources or knowledge of complicated mathematical operations. Given sufficient number of shares, a layman can reconstruct the secret graph. No mathematical knowledge is assumed, just the intuitive ability to pick common elements from two given sets. Such a feature is found in Visual cryptography (\cite{visual}). However, Visual cryptography requires shares to be printed on transparencies which need to be aligned perfectly for reconstruction.
\item In certain cases, our method has the advantage of visual decoding. i.e., we do not require any calculating devices or computers. It is notable that this is achieved while retaining information theoretic security.
\item The same share may be used for sharing two different secrets (may or may not be from the same domain). In such a scheme, the elements corresponding to one secret become the items eliminated on intersection for the other secret and vice versa.
\item The scheme works basically by mixing additional irrelevant data with the secret to be shared. We can also implement the scheme in such a way that the irrelevant data mixed come from a variety of domains. For example, a share for a secret string may consist of a number of strings, numbers, matrices, sets and symbols. In that case, the parties with the secret shares do not even know what kind of object is being shared. Only during the time of reconstruction, this information is revealed. This is implemented identical to the sharing of a secret set discussed in Section \ref{section:set}.
\item In Shamir's scheme, even if there occurs a $1$ bit error somewhere in the $k$ shares used for reconstruction, it fails. However, our scheme designed with inherent redundancy is robust enough to work in conditions of error worse than the one mentioned.

\end{enumerate}

\section{Drawbacks and possible solutions}
\begin{enumerate}
\item Since our algorithm does not take into account specific attributes that determine how a graph from a particular domain would look, there could be some methods to attack it. For example, when every node of the graph represents a distinct amino
acid residue in a protein and has the residue type as its label \cite{prote}, looking at a share, a domain expert might be able to say that such a bond between molecules is not possible and hence the edge could be removed. The secret, in these cases would be graphs with upto $300$ nodes \cite{prolen}. Thus, this could be seen as a way to attack the scheme in that domain. Since, shares are created at random, we do not take into account the feasibility of a particular edge in the underlying domain. However, consulting domain knowledge while constructing shares can solve this problem although then, creation of shares needs to be carried out with greater care.
\item An ideal $(k, n)$ secret sharing scheme is one in which the information gained from more than one share or even $k-1$ shares must be equal to that gained from one share. Although our scheme satisfies this for $(2, n)$ trivially, for $(3, n)$ schemes, this is not achieved. However, we can choose the amount of extra information to be added in such a way that the effort required to isolate the secret out of information gathered by pooling together $(k-1)$ shares is still greater than that of the brute force approach.
\end{enumerate}

\section{Conclusion}
We conclude by observing that existing secret sharing algorithms implicitly assume that the secret to be shared is encoded in to a number (with the exception of Visual cryptography). Such an encoding is not only difficult for sharing graphs but also requires public knowledge of the encoding scheme. Our method of sharing graphs works directly with graphs and encoding into numbers is eliminated. Further, our method has the advantage of visual reconstruction of the secret for small graphs since we use graph intersection unlike conventional techniques which rely on modular arithmetic and polynomial interpolation. What we propose could be thought of as a \emph{pencil and paper} secret sharing scheme which could be used even by a layman with no mathematical expertise nonetheless. As an example, $8$-length passwords which are very common, can be conveniently shared using our approach. Computation-wise, we make use of two basic operations namely, assignment and checking for equality and our scheme outperforms Shamir's scheme while retaining perfect secrecy.

\section*{Acknowledgements}
We would like to extend our thanks to Jayaraj Poroor, Amrita Research Labs and Rajan S., Department of Mathematics for their invaluable suggestions.

\end{document}